# Accepted Manuscript

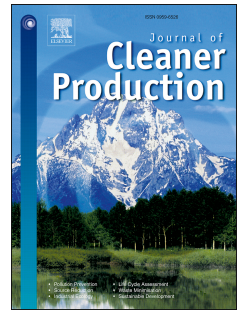

Beer-brewing powered by controlled hydrodynamic cavitation: Theory and real-scale experiments

Lorenzo Albanese, Rosaria Ciriminna, Francesco Meneguzzo, Mario Pagliaro

Please cite this article as: Albanese L, Ciriminna R, Meneguzzo F, Pagliaro M, Beer-brewing powered by controlled hydrodynamic cavitation: Theory and real-scale experiments, *Journal of Cleaner Production* (2016), doi: 10.1016/j.jclepro.2016.11.162.

This is a PDF file of an unedited manuscript that has been accepted for publication. As a service to our customers we are providing this early version of the manuscript. The manuscript will undergo copyediting, typesetting, and review of the resulting proof before it is published in its final form. Please note that during the production process errors may be discovered which could affect the content, and all legal disclaimers that apply to the journal pertain.

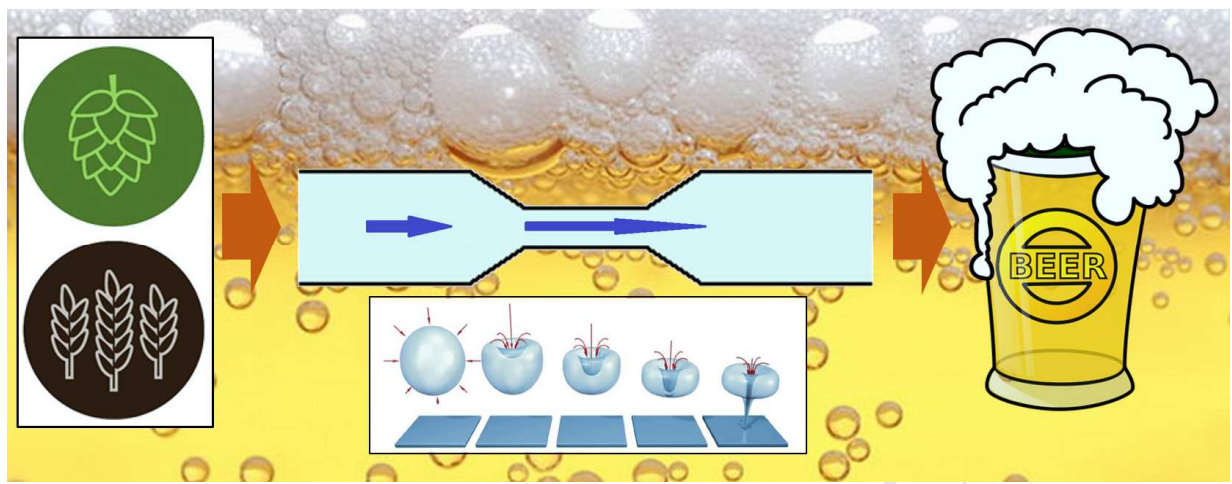

# Beer-brewing powered by controlled hydrodynamic cavitation: theory and real-scale experiments

Lorenzo Albanese[1]

Rosaria Ciriminna[2]

Francesco Meneguzzo[1*]

Mario Pagliaro[2*]

[1] Istituto di Biometeorologia, CNR, via G. Caproni 8, 50145 Firenze, Italy

[2] Istituto per lo Studio dei Materiali Nanostrutturati, CNR, via U. La Malfa 153, 90146 Palermo, Italy

**Corresponding author:** Dr. Francesco Meneguzzo, Istituto di Biometeorologia, CNR, via G. Caproni 8, 50145 Firenze, Italy. Tel. +39-392-9850002. Fax +39-055-308910. E-mail: f.meneguzzo@ibimet.cnr.it

**Corresponding author:** Dr. Mario Pagliaro, Istituto per lo Studio dei Materiali Nanostrutturati, CNR, via U. La Malfa 153, 90146 Palermo, Italy. Tel. +39 091 6809370. Fax +39 091 6809399. E-mail: mario.pagliaro@cnr.it

**Abstract**

The basic beer-brewing industrial practices have barely changed over time. While well proven and stable, they have been refractory to substantial innovation. Technologies harnessing hydrodynamic cavitation have emerged since the 1990s' in different technical fields including the processing of liquid foods, bringing in advantages such as acceleration of extraction processes, disinfection and energy efficiency. Nevertheless, so far beer-brewing processes were not investigated. The impacts of controlled hydrodynamic cavitation, managed by means of a dedicated unit on a real microbrewery scale (230 L), on the beer-brewing processes is the subject of this paper. The physico-chemical features of the obtained products, analyzed by means of professional instruments, were compared with both literature data and data from the outcomes of a traditional equipment. Traditional processes such as dry milling of malts and wort boiling becoming entirely unnecessary, dramatic reduction of saccharification temperature, acceleration and increase of starch extraction efficiency, relevant energy saving, while retaining safety, reliability, scalability, virtually universal application to any brewing recipe, beer quality, were the most relevant experimental results. The impacts of these findings are potentially far reaching, beer being the worldwide most widely consumed alcoholic beverage, therefore highly relevant to health, environment, the economy and even to local identities.

**Keywords:** beer; brewing; energy efficiency; hops; hydrodynamic cavitation; malt.

## 1. Introduction

Likely produced as early as of 7,000 BC (Pires and Brányik, 2015), beer is the alcoholic beverage most widely consumed around the world, made out of water, malt or grains, hops and yeasts as basic ingredients, which – along with the brewing technology – have barely changed over time, except for the diversification of ingredients occurred in the recent centuries (Ambrosi et al., 2014). In 2014, the global beer production, led by China, the United States and Brazil, amounted to about 1.96 billion liters, up from 1.3 billion liters in 1998. While the market is truly international, however, beer production is intertwined with notions of national identity, culture and pride (Stack et al., 2016). .

Being firmly established since so long ago, the main brewing steps have been extensively illustrated and reviewed in the existing literature (Pires and Brányik, 2015), along with the respective microbiological processes (Bokulich and Bamforth, 2013). A recent and up to date scheme of production processes at a craft brewery can be found in Figure 1 of the study by Kubule and coauthors (Kubule et al., 2016).

This study critically investigates how controlled hydrodynamic cavitation (HC) assisted brewing can help improving few stages of the beer production process, in terms of production time and yield, energy efficiency, beer quality, while retaining or improving manageability, stability, repeatability, reliability and scalability. The distinct effects of HC-assisted brewing are identified by comparing the respective outcomes with published data as well as with data from conventional equipment and processes. In particular, the investigation – carried out at a real microbrewery scale – included key brewing steps such as malt milling, mashing, hopping and boiling.

The idea to introduce HC in the brewing process arose from its few established and attractive features such as straightforward scalability, stability and localized high density energy release to the liquid medium, including liquid foods (Gogate, 2011; Gogate and Pandit, 2001).

Indeed, applications to liquid food processing involving the use of cavitation processes date back to the early 1990s, when cavitation was mostly induced by ultrasound irradiation of small liquid volumes, also known as acoustic cavitation (AC) or sonocavitation. However, hydrodynamic cavitation was later shown to far outperform acoustic cavitation as to the respective typical energy efficiency, *i.e.* the ratio of the power dissipated in the liquid to the electric power supplied to the system, the relative advantage growing with size (Gogate et al., 2001; Langone et al., 2015). The same holds for the cavitational yield, defined as the actual net production of desired products per unit supplied electrical energy (Gogate et al., 2001; Gogate and Pandit, 2005). Furthermore, compared to AC cavitators, HC reactors have revealed more suitable for industrial applications due to wider area of cavitation, much lower equipment cost and more straightforward scalability.

Referring to another recent study by the authors for a deeper discussion on the principles, history and emerging industrial applications of the controlled hydrodynamic cavitation (Ciriminna et al., 2016a), few key issues will be recalled in the following.

Constrictions and nozzles, resulting in acceleration and local depressurization, alter the flow geometry. If the pressure falls below the boiling point, water vaporizes and vapor bubbles are generated. For the purposes of the present study, different HC regimes are practically identified according to the values assumed by a single dimensionless parameter, *i.e.* the cavitation number – indicated as CN or $\sigma$ in the following – derived from Bernoulli's equation and shown in its simplest form in equation 1.

$$\sigma = (P_0 - P_v) / (0.5 \cdot \rho \cdot u^2) \qquad (1)$$

where $P_0$ ($Nm^{-2}$) is the average pressure measured downstream of a cavitation reactor, such as a Venturi tube or an orifice plate, where cavitation bubbles collapse, $P_v$ ($Nm^{-2}$) is the liquid vapor pressure (a function of the average temperature for any given liquid), $\rho$ ($kgm^{-3}$) is the liquid density, and $u$ ($ms^{-1}$) is the flow velocity measured through the nozzle of the cavitation reactor. Such simplified representation was recently supported by scientists in Japan, who showed that the control of the downstream pressure ($P_0$) allows describing the most relevant and desired features of hydrodynamic cavitation (Soyama and Hoshino, 2016).

Few years before, Gogate and coworkers identified three intervals in the range of values assumed by the cavitation number, corresponding to broad cavitational regimes (Bagal and Gogate, 2014; Gogate, 2002), as well represented in Scheme 1 in the above-mentioned study of the authors (Ciriminna et al., 2016a). Ideally, without impurities and dissolved gases, cavities would be generated for values of $\sigma$ below 1, the latter to be meant as an average value, depending on the specific setup. Below a threshold around 0.1 on average, cavities are no more able to collapse and the HC regime turns to chocked cavitation or supercavitation. For $\sigma$ growing over 1, lesser and lesser cavities are generated, while their collapse becomes ever more violent. For the scope of this study, only the developed cavitation with $0.1 < \sigma < 1$ will be considered.

Quite recently, the validity of the very concept of cavitation number was questioned by Šarc and coauthors (Ciriminna et al., 2016a; Šarc et al., 2017), who issued a comprehensive set of suggestions and recommendations aimed at improving the understanding and repeatability of HC processes and experiments, which will be taken into account in the following, particularly in

Section 2.1.

Among controlled hydrodynamic cavitation devices, where HC regimes can be tuned over a wide range while avoiding any damage to the equipment, non-stationary reactors, such as rotational generators, were successfully developed and applied. Nevertheless, Venturi-type stationary reactors still appear as the most appealing candidates for industrial-scale applications due to their superior cheapness as well as intrinsic ease of construction, scaling and replicability. Venturi-tube reactors are as well usually preferred over orifice plates running the risk of obstruction from solid particles and other viscous substances such as found in brewing applications (Albanese et al., 2015).

## 2. Materials and methods

### 2.1 Brewing units

A dedicated equipment was built from known or commonly available commercial components, in order to investigate the effects of hydrodynamic cavitation processes upon few brewing steps and respective outputs.

Figure 1 shows the experimental installation, including a closed hydraulic loop with total volume capacity around 230 L, powered by a centrifugal pump (Lowara, Vicenza, Italy, model ESHE 50-160/75 with 7.5 kW nominal mechanical power) with open impeller 0.174 m in diameter. Rotation speed was set around 2900 rpm. As shown in the manufacturer's technical documentation at page 48 ("Serie e-SH (in Italian)," 2016), the maximum pressure and volumetric flow rate were around 4 atm and 1,500 Lmin$^{-1}$, respectively.

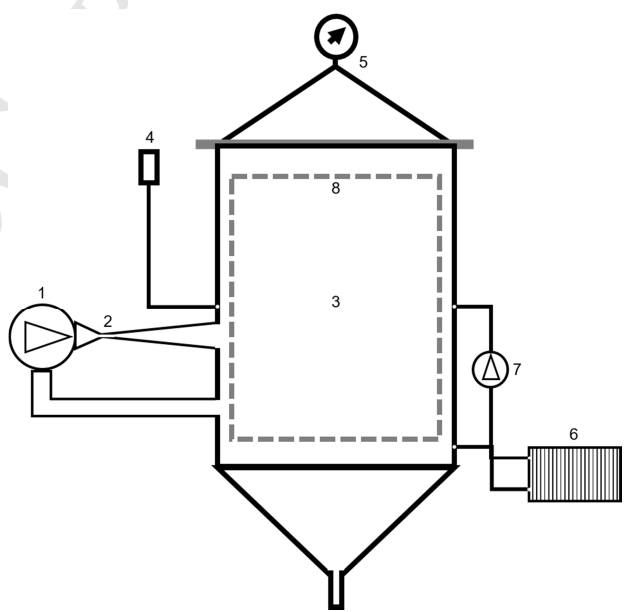

**Figure 1**. Simplified scheme of the experimental HC-based installation. 1 – centrifugal pump, 2 – HC reactor, 3 – main vessel, 4 – pressure release valve, 5 – cover and pressure gauge, 6 – heat exchanger, 7 – circulation pump, 8 – malts caging vessel. Other components are commonly used in state-of-the-art hydraulic constructions.

Any surface in contact with the wort was crafted in food-quality stainless steel (AISI 304), with 2 mm minimum thickness. The circulating liquid (wort) can be exposed to the atmospheric pressure or to a given average pressure limited by a tunable pressure release valve. Such valve was preferred over an expansion tank in order to avoid wort contamination by substances accumulated in the tank during successive tests, while performing the same task, i.e. tuning the cavitation intensity through the $P_0$ term in equation (1).

Volumetric liquid heating occurs during circulation due to the conversion of impeller's mechanical energy into thermal energy, particularly downstream the cavitation reactor nested into the hydraulic loop, due to the vigorous internal friction associated with the cavitation process. As a closed hydraulic circuit, no change of potential energy is involved, therefore all the mechanical energy turns into heat.

A Venturi tube, with the same geometry described in detail in Figure 2(B) of a previous study by the authors (Albanese et al., 2015), is used as the cavitation reactor and preferred over an orifice plate since it was observed that orifices are quickly obstructed by the circulating solid particles. Moreover, a smaller circulation pump (Rover pompe, Padova, Italy, model NOVAX 20 B, power 340 W, working temperature up to 95°C, capacity up to 28 L min$^{-1}$), drives a secondary recirculation loop through an ordinary plate heat exchanger (20 stainless steel plates, each with a 0.043 m$^2$ surface area), allowing for isothermal stages when required in the course of the production process, depending on specific brewing recipes. The latter pump was used after hopping, both to cool the wort and to convey it to the fermenters.

The design allows for upscaling of a single installation unit up to the order of 10,000 L, for housing further pumps and cavitation reactors, and for straightforward integration of isolated components, such as pumps and HC reactors, into existing brewing and fermentation plants of virtually any size. The advantages of parallel arrangement, obvious for pumps, were originally found in a previous study (Maeng et al., 2010).

Onboard sensors include a pressure gauge and few digital thermometers (not shown), hydraulic pressure and temperature being the only relevant physical parameters monitored and actually used to manage the brewing processes as well as to characterize the cavitation regimes.

It is important to note that, following recent recommendations (Šarc et al., 2017; Soyama and Hoshino, 2016), the wort pressure $P_0$ in equation (1) was measured far downstream the nozzle of the Venturi reactor, as shown by the location of the pressure gauge in Figure 1. Following the same recommendations, the flow velocity $u$ through the nozzle of the same cavitation reactor was considered. In particular, no flowmeter having been available, the flow capacity ($m^3 s^{-1}$) was estimated on the basis of the pump's own characteristic curves – also shown in the manufacturer's technical documentation at page 48 ("Serie e-SH (in Italian)," 2016) – describing the relationship between capacity, head (m) and absorbed power (W).

The power absorption by the centrifugal pump was measured by means of a three-phase digital power meter (IME, Milan, Italy, model D4-Pd, power resolution 1 W, energy resolution 10 Wh, accuracy according to the norm EN/IEC 62053-21, class 1). Once the water flow is set, the flow capacity was assessed every 5 min from the reading of the absorbed power, and the velocity $u$ through the nozzle was in turn computed from straightforward division of the flow capacity by the nozzle's section.

While the liquid vapor pressure $P_v$ was computed on the basis of the temperature readings, the water density ($10^3 kg/m^3$) was used throughout all calculations, leading to a small underestimation of the cavitation number, since malt and starch are less dense than water.

While conceding that not all the suggestions advanced by Šarc and coauthors were met by the above explained methodology (Šarc et al., 2017), let alone a so far unavailable theoretical framework for HC science and technology, the supplied information should allow at least the repeatability of the experiments carried out with the same setup including pump, cavitation reactor and size of the brewing device.

While this installation was designed to perform the mashing and hopping stages of brewing, and fermentation was generally performed in common 200 L stainless steel cylinder-conical fermenters after receiving the wort from the main unit, few tests were performed without wort removal, using the installation shown in Figure 1 as the fermenter.

The malts caging vessel represented in Figure 1 was used in few experiments – also referred to in the following with the common name of brew in the bag (BIAB) – to better isolate the effects of hydrodynamic cavitation on the saccharification and starch extraction efficiency. In this case, the malts are not allowed to circulate, being caged in a cylindrical vessel made up of a stainless steel fine grid with a perforated pipe arranged along the vessel axis, connected to the same external pump used for thermal stabilization. The forced wort circulation is meant to boost the extraction of starch and enzymes from the malts. In BIAB tests, malt milling before mashing was required and performed by means of a small semiautomatic stainless steel roller mill. On the contrary, hops – being pitched after the removal of the cylindrical vessel – were allowed to circulate.

While in the BIAB tests the removal of spent cavitating malts after completing the mashing stage is straightforward , in the other experiments a time consuming mechanical filtering and gravity separation was needed to remove most solid particles. In industrial applications, the use of a commercial centrifuge, e.g. a decanter centrifuge for soft sediment, is recommended. Water, before being conveyed to the production unit, is passed through a mechanical filter made up of a 20 µm polypropylene wire to remove solids particles down to 50 µm in size. An active carbon filter reduces chlorine concentration, attenuates odors and flavors, and removes other impurities down to 70 µm in size. The pH is usually lowered from about 7 to about 5.5 by adding 80 wt% lactic acid (70-80 mL).

For comparison purposes, traditional brewing was performed by means of a Braumeister (Ofterdingen, Germany) model 50 L brewer, equipped with a cooling serpentine (model Speidel stainless steel wort chiller for 50-liter Braumeister) and fully automatic brewing control (temperature, time and recirculation pumps).

Finally, after fermentation, bottling was performed via an ordinary depression pump (Tenco, Genova, Italy, model Enolmatic, with capacity around 200 bottles per hour).

### 2.2 Measurement instruments and methods

Along with thermometer and manometer sensors onboard the main production unit, few specialized off-line instruments were used to measure the chemical and physiological properties of wort and beer.

The acidity was measured by means of pH-meter (Hanna Instruments, Padova, Italy, model HI 99151) with automatic pH calibration and temperature compensation. The sugar concentration in the wort during mashing and before fermentation was measured in Brix percentage degrees by means of a refractometer (Hanna Instruments, Padova, Italy, model HI 96811, scaled from 0% to 50% Brix, resolution 0.1%, precision ±0.2% in the 0-80°C temperature range, and automatic temperature compensation in the 0-40°C range). Brix readings were then converted to starch extraction efficiency (Bohačenko et al., 2006).

Physico-chemical and physiological parameters of fermenting wort and finished beer were measured by means of a 6-channel photometric device (CDR, Firenze, Italy, model BeerLab Touch). In particular: fermentable sugars (0.1 to 150 g/L of maltose, resolution 0.01 g/L), alcohol content (0-10% in volume, resolution 0.1%), bitterness on the International Bitterness Unit (IBU) scale (5 to 100, resolution 0.1) (Lajçi et al., 2013), color on the European Brewery Convention (EBC) scale (1 to 100, resolution 1) and on the Standard Reference Method (SRM) scale (0.5 to 50, resolution 0.1). All reagents were of analytical grade.

The electricity consumed by the centrifugal pump used in the HC device shown in Figure 1 was measured by means of the same commercial digital meter used to measure its three-phase power absorption, mentioned in Section 2.1. The electricity consumed by the Braumeister model 50-liters traditional brewing device was measured by means of a mono-phase -phase digital energy meter (PeakTech, Ahrensburg, Germany, model 9035, power resolution 1 W, energy resolution 100 Wh, accuracy 0.5%).

## 2.3 Brewing ingredients

Pilsner or Pale were used as the base barley malts in all the performed tests, along with smaller fractions of Cara Pils, Cara Hell and Weizen, the latter ones supporting body, flavor, aroma and foam stability of the finished beer.

Among the hops, different combinations of pelletized German Perle, Saaz and German Hersbrucker were used. In the course of few tests, candied brown sugar was added to the wort before fermentation, while regular white sugar was added to the fermented wort before bottling and maturation. Finally, fermentation was activated by means of the dry yeast strain Safale US-05, requiring temperature between 15°C and 24°C and maximum alcohol content 9.2%, used in any test in the identical proportion of 67 g per 100 L.

## 2.4 Production tests

Table 1 summarizes few basic features of the performed brewing tests. It should be noted that no simple sugar was added during the mashing stage in any test. Two tests (CO2 and CO3) were stopped before fermentation. Three tests (C8, C9 and C10) were performed under exactly the same operational conditions (*i.e.*, same ingredients) to study the effects of structural changes to the malts caging vessel. Five tests (IBU1-IBU5) were carried out without malts to study only the utilization of hops' α-acids. Finally, two tests (B1 and B2) were performed with a traditional equipment.

**Table 1**. Beer production tests, ingredients and conditions.

| Test ID | Production unit[a)] | Net volume (L) | Malt | Cavitating malts | Hops[b)] | Added sugars[d)] | Fermentation[e)] |
|---|---|---|---|---|---|---|---|
| CO1 | 1(A) | 186 | Pilsner 25 kg<br>Cara Pils 1.6 kg<br>Cara Hell 2.6 kg<br>Weizen 2 kg | Yes | Perle 0.1 kg<br>Hers[c)] 0.3 kg<br>Saaz 0.2 kg | W 0.96 kg (bot) | Standard |
| C2 | 1(B) | 170 | Pilsner 25 kg<br>Cara Pils 1.6 kg<br>Cara Hell 2.6 kg<br>Weizen 2 kg | No | Perle 0.1 kg<br>Hers 0.4 kg<br>Saaz 0.1 kg | W 0.96 kg (bot) | Standard |
| C5 | 1(B) | 170 | Pilsner 25 kg<br>Cara Pils 3.6 kg<br>Cara Hell 2.6 kg | No | Perle 0.1 kg<br>Hers 0.3 kg<br>Saaz 0.2 kg | W 10 kg (fer)<br>W 0.96 kg (bot) | Standard |
| C6 | 1(B) | 170 | Pilsner 25 kg<br>Cara Pils 3.6 kg<br>Cara Hell 2.6 kg | No | Perle 0.3 kg<br>Hers 0.4 kg<br>Saaz 0.2 kg | W 8 kg (fer)<br>W 0.96 kg (bot) | Standard |
| B1 | B-50 | 50 | Pilsner 6.25 kg<br>Cara Pils 0.9 kg<br>Cara Hell 0.65 kg | No | Perle 0.025 kg<br>Hers 0.075 kg<br>Saaz 0.05 kg | W 0.042 kg (bot) | Standard |
| CO2 | 1(A) | 186 | Pilsner 25 kg<br>Cara Pils 3.6 kg<br>Cara Hell 2.6 kg | Yes | | | |
| CO3 | 1(A) | 186 | Pilsner 25 kg<br>Cara Pils 6.2 kg | Yes | Neither hopping nor fermentation performed | | |
| C7 | 1(B) | 170 | Pilsner 28.5 kg<br>Cara Pils 2.5 kg | No | Perle 0.6 kg<br>Saaz 0.5 kg | W 0.84 kg (bot) | Standard |
| B2 | B-50 | 50 | Pilsner 9.2 kg<br>Cara Pils 0.81 kg | No | Perle 0.194 kg<br>Saaz 0.162 kg | W 0.07 kg (bot) | Standard |
| C8 | 1(B) | 170 | Pale 26 kg<br>Cara Pils 3 kg | No | Perle 0.2 kg<br>Hers 0.1 kg | B 1.0 kg (fer)<br>W 0.96 kg (bot) | Installation |
| C9 | 1(B) | 170 | Pale 26 kg<br>Cara Pils 3 kg | No | Perle 0.2 kg<br>Hers 0.1 kg | B 1.0 kg (fer)<br>W 0.96 kg (bot) | Installation |
| C10 | 1(B) | 170 | Pale 26 kg<br>Cara Pils 3 kg | No | Perle 0.2 kg<br>Hers 0.1 kg | B 1.0 kg (fer)<br>W 0.96 kg (bot) | Installation |
| Next tests were designed to study the utilization of hops' α-acids | | | | | | | |
| IBU1 | 1(A) | 225 | | | Perle 0.3 kg | | |
| IBU2 | 1(A) | 225 | | | Perle 0.3 kg | No fermentation performed | |
| IBU3 | 1(A) | 228 | | | Perle 0.3 kg | | |

| IBU4 | B-50 | 55 | Perle 0.076 kg |
|------|------|-----|----------------|
| IBU5 | 1(A) | 228 | Perle 0.3 kg |

[a]  1(A) = installation shown in Figure 1. 1(B) = variant with malts caged in a cylindrical vessel (BIAB). B-50 = Braumeister model 50-liters.

[b]  Hops cavitating in any test.

[c]  Hers = Hallertau Hersbrucker hop.

[d]  W = simple white sugar. B = candied brown sugar. bot = before bottling. fer = before fermentation.

[e]  Standard = fermentation in cylinder-conical vessel 200 L. Installation = fermentation performed into the experimental installation.

## 3. Theoretical background

Physical phenomena associated with the collapse of cavitation bubbles were described by Yasui and co-authors in Japan while studying single-bubble and multi-bubble sonoluminescence, showing that the temperature and pressure inside a collapsing bubble increases dramatically up to 5,000-10,000 K and 300 atm, respectively, due to work done by the liquid to the shrinking bubble (Yasui et al., 2004). The same authors predicted as well the formation of hydroxyl radicals ($\cdot$OH) by developed HC, as a result of water splitting, whenever the internal temperatures of the collapsing bubbles grow over 2,500 K. Yet, few years passed until the splitting processes and formation of those powerful oxidants (oxidation potential 2.80 eV) were theoretically and experimentally demonstrated by several independent authors working with developed cavitation regimes ($0.1 < \sigma < 0.25$) (Batoeva et al., 2011; Rajoriya et al., 2016; Soyama and Hoshino, 2016).

As a result of the peculiar physical and chemical effects of controlled hydrodynamic cavitation, its applications to water and wastewater remediation – degradation of organic pollutants, reduction and solubilization of the overall organic loads, and disinfection from pathogen microorganisms – have received growing attention both in the scientific literature and industrial practice, for which recent comprehensive reviews are available (Dindar, 2016; Dular et al., 2016; Tao et al., 2016).

A potential issue concerns oxidation processes that are generally undesirable within the field of liquid foods processing (Ngadi et al., 2012). However, within the range of HC regimes used in the field of food applications, oxidation processes have been shown to play quite a marginal role in comparison to straightforward mechanical effects generated by the collapse of cavitation bubbles (Yusaf and Al-Juboori, 2014). Indeed, only the use of specific additives such as hydrogen peroxide allows achieving the needed extent of organics oxidation in applications such as water disinfection and remediation (Ciriminna et al., 2016b).

No other harmful side effect were observed to prevent HC applications to liquid foods processing in assistance or replacement of traditional techniques, such as those aimed to pasteurization and homogenization, moreover allowing to perform volumetric heating (via conversion of mechanical energy into heat) which offers the advantage to avoid thermal gradients and consequently caramelization hazards (Albanese et al., 2015).

The process of saccharification, *i.e.* the conversion of malt starch to simple fermentable sugars (glucose, maltose, and maltotriose) by malt enzymes such as α- and β-amylases (Hager et al., 2014), shows a significant acceleration and intensification when ultrasonically irradiated, as long as the irradiation power does not exceed a certain threshold, beyond which the enzymes are damaged and inactivated (Knorr et al., 2011; Sinisterra, 1992). Such evidence, linked to the increase of the mass transfer rate from reactants to enzymes, was attributed to the increase of the permeability of cellular membranes, and to the transformation of the liquid boundary layer surrounding the cellular walls (Knorr et al., 2011). More recently, beneficial effects from AC were observed in the extraction of polyphenols, simple sugars and mineral elements in apple juices (Abid et al., 2014, 2013), and in blueberry juices (He et al., 2016).

Most of results from the research concerning the application of devices comprising HC reactors to the technical field of liquid foods, arisen during the last two decades, involved laboratory-scale liquid volumes, and foods other than beer, such as fruit juices (Milly et al., 2008, 2007), milk and yogurt (Sfakianakis and Tzia, 2014). Generally, the aim was pasteurization, namely to extend the shelf life, while preserving or improving the organoleptic and nutritional qualities (Albanese et al., 2015; Gogate, 2011; Ngadi et al., 2012).

When applied to the primary fermentation stage of few food liquids, low power sonocavitation was shown to increase the ethanol concentration, sometimes reducing the fermentation time (*e.g.*, to 64% of the time needed for untreated samples, in the case of beer), due to the removal of dissolved carbon dioxide by the AC-induced cavitation processes (Matsuura et al., 1994). Evidence was shown that the ultrasound irradiation within a certain power range lowered the overall acidity, increased the size and density of yeast cells, thus boosting their fermentation efficiency, as well as boosted the amino acids utilization during fermentation (Knorr et al., 2011; Matsuura et al., 1994). .

Recently, a sample of wort beer after yeast pitching was treated in an ultrasonic bath of given frequency and variable power (Choi et al., 2015). The increase of ethanol production during the subsequent fermentation increased with sonication power up to a peak around 13%, and decreased with greater power. Around the same peak conditions, an acceleration of the sugar assimilation was attributed to the increased permeability of yeasts' cellular membranes and consequently of the mass transfer, as well as an acceleration of the free amino-nitrogen (FAN) utilization was observed. The latter aspect was deemed to affect the beer's organoleptic and physiological properties likely by means of the enhanced assimilation of proline. Physico-chemical and sensorial properties of the finished product were shown to be unaffected by the AC treatment, at least below the peak power.

Although achieved by means of ultrasound-assisted cavitation in laboratory-scale experiments, those early results provided a useful guidance for industrial scale HC applications.

Further recent works have shown the beneficial application of hydraulic pressures exceeding the atmospheric value to the mashing processes, increasing the yield of fermentable sugars by means of the starch gelatinization along with enzymes extraction and activation, up to 20 atm (Ahmed et al., 2016; Choi et al., 2016). This is relevant in view of the localized wide pressure waves generated because of the collapse of cavitational bubbles.

Quite recently, a combined AC and thermal application showed to exceed the purely thermal treatment in the inactivation efficiency of *Saccharomyces cerevisiae* ascospores in fermenting beer wort, around the temperature of 60°C (Milani et al., 2016). The flaw represented by the low energy efficiency, due to the poor performance of the AC process, can be eliminated when HC processes replace energy inefficient sonocavitation (Albanese et al., 2015).

As a further advantage of controlled hydrocavitation applied to liquid foods, the focusing of the heating source in the cavitation active volume reduces thermal dissipation through pipe sidewalls, section changes, curves and other discontinuities in the hydraulic piping, leading to energy savings that grow with the operating temperature (Baurov et al., 2014).

To the authors' best knowledge, an only published study concerned the application of HC techniques to the fundamental stages of beer production, by means of a rotating-pulsating reactor immersed in the liquid (Safonova et al., 2015). The declared results, limited to laboratory-scale volumes (1.12 liters), indicated a very relevant reduction of the mashing time, down to just 10-15 min, as well as a shortening of the fermentation time by 1.5 days, in comparison to traditional methods, along with the improvement of physiological and biochemical properties of the culture, without damaging the quality of the finished product.

## 4. Results

HC-assisted brewing can significantly affect the process in every step it is applied to. In the following, both the potential effects, derived from literature and past experience, and the observed ones are discussed in detail.

### 4.1 Water and wort disinfection

Water conveyed to the brewing processes needs to be chemically and microbiologically pure. Before mashing, water is disinfected and purified in order to remove microbial pathogens as well as other possible organic and inorganic contaminants. Adjusting water acidity can be performed at this stage.

While not experimentally investigated in this study, the recent literature hints the feasibility of water remediation directly linked to the application of HC processes to brewing. For example, Dular and co-authors Slovenia recently offered a comprehensive review, along with their own experimental tests, of controlled HC as an emerging economically and technically feasible alternative to degrade and eventually mineralize harmful chemical and biological pollutants, as well as to inactivate pathogen microorganisms, via advanced oxidation processes and mechanical effects (Dular et al., 2016).

Beyond water, pathogen microorganisms can grow in malts and wort too, such as fungi, mycotoxins, *Bacillus* spp., *Clostridium*, wort spoilers such as Gram-negative enterobacteria, can be harmful to human health as well as to key brewing stages such as fermentation, hence detrimental to beer quality (Bokulich and Bamforth, 2013). Therefore, their inactivation is highly desirable and, in this respect, hydrodynamic cavitation appears as an appealing candidate.

**4.2 Dry milling**

Dry milling of malts, either barley or other grains, in traditional processes is aimed at upsizing the exchange interface between malts and water (wort) and thus the mass exchange rate of fermentable sugars and enzymes (Pires and Brányik, 2015). Hydrodynamic cavitation makes this step irrelevant, since shear forces and mechanical jets pulverize the grains down to less than 100 μm within few minutes, as checked during tests CO1, CO2 and CO3 described in Table 1 by means of a mechanical sieve. Figure 2 shows a sample of grains resulting from a traditional process and after HC application.

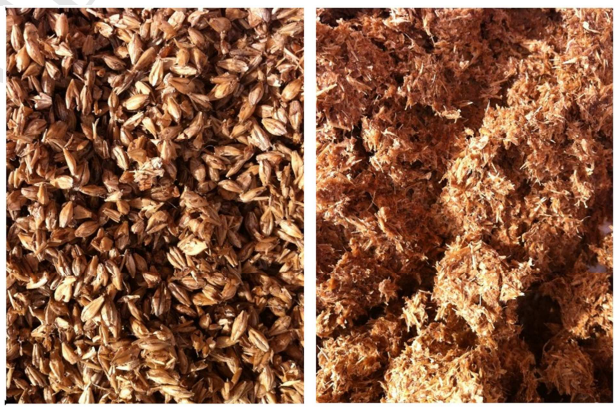

**Figure 2**. Malt particles after a traditional brewing process (*left*) and after hydrodynamic cavitational processing (*right*).

Besides offering the typical advantages of wet milling (Szwajgier, 2011), the new process affords a further increase of the starch extraction rate as well as of its overall concentration in the wort, as will be shown in Section 4.4. Moreover, the above mentioned pulverization hints to an increase of spent malts' solubilization and biodegradability, along with their methanogenic potential in the event they are sent to an anaerobic digester for energy and material recovery, thereby allowing a significant reduction of the biomass waste to be disposed of (Lee and Han, 2013; Maeng et al., 2010; Patil et al., 2016).

Cleaning and sanitization of brewing installations and the production environment, while fundamental to productivity and food safety, are energy and water intensive. Wastewater volumes generated by breweries have been estimated in the range of 4 to 11 liters per liter of finished beer, with small workshops on the upper limit. Moreover, such wastewater carries an high content of dissolved organic substances, which originate a chemical oxygen demand (COD) between 2000 and 6000 mg/L, namely so much significant that often wastewater is treated on site prior to discharge (Amienyo and Azapagic, 2016; Pettigrew et al., 2015). In certain plants, wastewater and spent malts are transferred to anaerobic digesters for biogas production. In this respect, based on the arguments advanced in Section 4.1 and in this Section, HC-assisted brewing can help both sanitizing and disinfecting water and wort, thereby potentially reducing the cleaning water needs, and solubilizing the spent malts, thereby enhancing their subsequent biological degradation.

### 4.3 Heating and saccharification

The mix of purified water and malts is heated to allow polysaccharides (starch) included in the malts to be hydrolyzed by malt enzymes to fermentable sugars and amino acids – in a process known as saccharification – that will be assimilated by yeast strains during fermentation. Such heating is performed either by electric resistance, flame or vapor as well as over the whole volume or part of it (the latter technique being known as decoction). Heating involves several sub-stages – including isothermal ones – in the temperature interval about 50°C to about 78°C, closely related to enzymes' action on wort's proteins (Pires and Brányik, 2015). Mechanical or hydraulic mixing, although gentle to minimize shear forces acting on the mash as well as increasing dissolved oxygen to prevent oxidative reactions, is needed in order to boost mass exchange, prevent caramelization and the occurrence of potentially harmful substances.

Besides its specific physico-chemical effects, hydrodynamic cavitation is a heating process too, as mentioned in Section 1 and Section 2.1, whose efficiency is limited only by the pumps' engines efficiency. As such, HC application to brewing prevents the use of any other power source or heating devices, as well as the volumetric nature of its heating process prevents the need for any other mechanical or hydraulic mixing.

Recalling the saccharification step as described in Section 1, Figure 3(A) shows that brewing in tests CO1 and CO3, performed with the setup sketched in Figure 1 achieves saccharification at 48°C and 37°C, respectively, against an average over 70°C for all other test. Remarkably, however, test CO2 carried out with an average pressure around 1.5 bar with much higher cavitation number, shows a saccharification temperature of 76°C. While further tests will be needed to identify the best cavitational regime, within the range spanned by the performed tests, Figure 3(B) shows that the CN values between 0.15 and 0.20, readily achievable at atmospheric pressure (as indeed they were in the experiments), seem recommendable, while higher CN regimes are likely to damage enzymes.

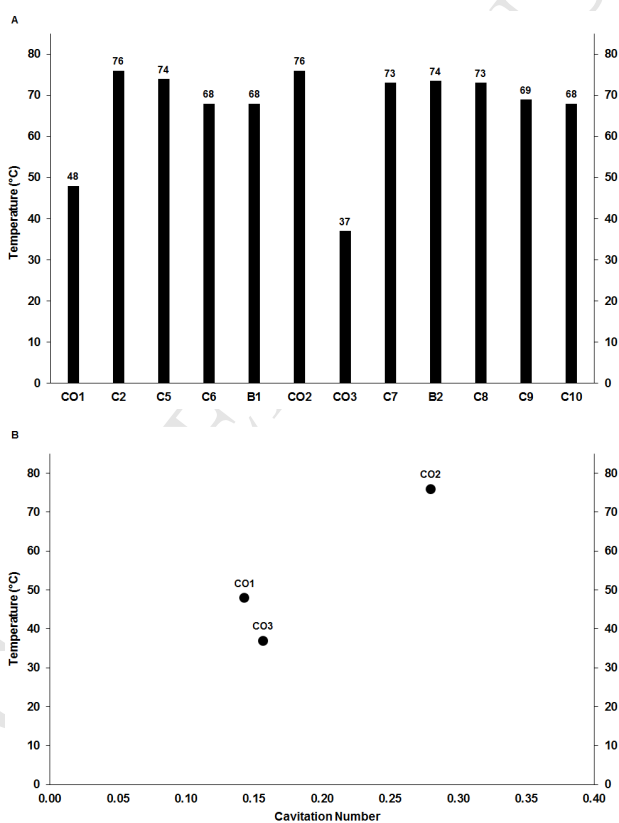

**Figure 3**. Saccharification temperature for all beer production tests (A), and saccharification temperature vs Cavitation Number for three selected production tests (B).

These results are in close agreement with previous experience gained with acoustic cavitation devices, as illustrated in Section 1. In particular, beyond the acceleration and intensification of the enzymatic activity under moderate cavitation conditions, the observed decreased efficiency under stronger hydrodynamic cavitation recalls the similar effect observed with acoustic cavitation and the related irradiation power beyond which the enzymes are damaged and inactivated (Knorr et al., 2011; Sinisterra, 1992). The interpretation according to which the increase of the mass transfer rate from reactants to enzymes was attributed to the increase of the permeability of cellular membranes, and to the transformation of the liquid boundary layer surrounding the cellular walls looks like to be confirmed by our experiments (Knorr et al., 2011).

In summary, it was observed that the activation temperature of enzymes aimed at transforming starch into simple sugars and amino acids drops by up to 35°C on average, thus shortening the time needed to saccharification. This evidence leads to the opportunity to start the circulation of the malts into an HC-assisted brewing device from the very beginning of the process, namely when water temperature is close to the room temperature.

### 4.4 Starch extraction efficiency

In traditional processes, malt removal and lautering, the latter in turn consisting of at least two steps, *i.e.* recirculation and sparging, are performed at a temperature around 78°C, before enzymes are fully deactivated (in particular, the enzyme alpha-amylase is still active), with lautering aimed at extracting the residual fermentable sugars. The whole brewing process carried out till that stage can take several hours and the starch extraction rate is generally limited below 80%, unless a more costly wet milling of malts is performed (Szwajgier, 2011).

Figure 4(A) shows the starch extraction efficiency over the considered tests. HC regimes established in tests CO1 and CO3 provided once again the best results (91% and 86%, respectively, of the starch extraction efficiency), thereby confirming the superiority of cavitating malts in the CN regime around 0.15, also in comparison to traditional brewing processes (tests B1 and B2).

Surprisingly enough, the overall starch extraction efficiency achieved with BIAB production test C10 (84%), approached the result obtained with test CO3. That result arose after repeated structural changes to the perforated pipe inside the vessel, as well as after strengthening the circulation in the same pipe, meant to increase the turbulence and, in turn, the extraction efficiency.

In other words, while cavitating malts with CN around 0.15 allows saccharification to occur at far lower temperatures than any other setup, as well as increasing the extraction efficiency by 20% to 30% over the average of any other setup, a careful management of the turbulence in the malt vessel may also provide very good results in terms of extraction efficiency. However, Figure 4(B) shows that the peak starch extraction efficiency achieved in test C10 came at the expense of process time and specific energy consumption, the latter more than 3 times higher than in both tests CO1 and CO3.

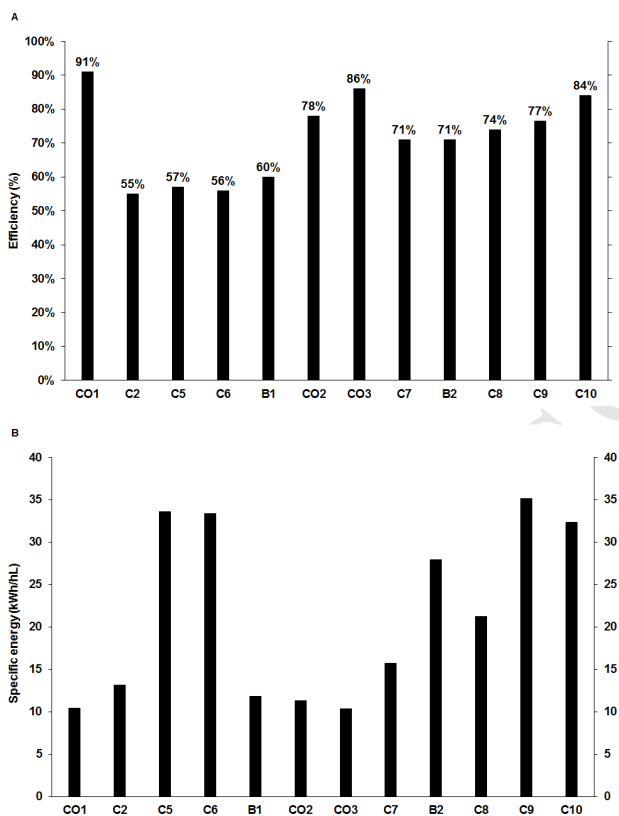

**Figure 4**. Starch extraction efficiency (A) and specific energy consumption at peak starch extraction rate (B) for all the beer production tests.

Comparing the production test CO1 with test B2 (performed by means of a traditional equipment), the latter provided a much lower starch extraction efficiency of 71%, as well as an energy consumption at its peak extraction rate about 28 kWh/hL, i.e. approximately 2.7 times higher than in test CO1 (10.4 kWh/hL).

As shown in Section 4.2, the application of HC processes pulverizes the malts, if the latter are allowed to circulate as in tests CO1, CO2 and CO3 (Table 1), suggesting that no significant starch and sugar concentration remains trapped into the grains. The results shown in Figure 4(A) look like to confirm the argument.

### 4.5 Hopping and wort boiling

In traditional brewing processes, after the removal of malts around 78°C (mashing-out), the wort is heated up to the boiling point (usually around 102°C) and boiled for some time. The boiling stage aims at performing several substantial tasks. Among them, extracting hops' α-acids and isomerizing them into soluble iso-α-acids which supply the desired bitterness, building the desired aromatic compounds, further inactivating enzymes, sterilizing and pasteurizing. Moreover, protein precipitation (hot-break) and formation of colloids contributing to foam and haze stability are further relevant objectives of boiling.

Lasting at least 60 min and usually 90 to 120 min (Pires and Brányik, 2015), boiling is usually the most energetically demanding among traditional brewing stages (Muster-Slawitsch et al., 2011; Olajire, 2012).

In traditional brewing processes, hops are pitched at temperatures not lower than 90°C, *i.e.* the temperature above which the extraction and the isomerization of hops' α-acids begins, growing slowly up to the boiling point. Nevertheless, the overall concentration of α-acids is limited by their own degradation, the latter in turn positively correlated with both temperature and process time. Such first-order kinetic process sets limits on the hops residence time in the wort, which is a main reason why the hops are generally pitched during the boiling, as well as on the boiling time itself and on the net utilization of α-acids, the latter topping in the most favorable cases at no more than 50% after about 90 min of boiling (Lajçi et al., 2013; Malowicki and Shellhammer, 2005).

Another important task must be faced in traditional brewing, namely the removal of undesired volatile aromatic compounds such as $(CH_3)_2S$ (dimethyl sulphide, DMS). DMS in wort and beer originates from known precursors in barley malts (Scheuren et al., 2014), its concentration as a dissolved gas increasing while the wort heats up due to the negative correlation between DMS volatility and wort temperature, thus complicating its removal (H. Scheuren et al., 2015). Unless suitable gas extractors are used, which are costly, energy-demanding and possibly leading to oxidation issues, wort boiling is the only process allowing effective evaporation of the dissolved DMS, as well as its concentration to fall below the odor threshold, at around 50 to 61 μg/L (Hans Scheuren et al., 2015; Scheuren et al., 2016).

HC-assisted extraction and isomerization of hops α-acids was quantitatively assessed by means of the so called utilization factor (Malowicki and Shellhammer, 2005). Figure 5 shows the outcomes of α-acids utilization in the course of the last five tests listed in Table 1 (IBU1-IBU5) designed to study these processes using only Perle hop, the latter having a mass fraction of α-acids at the level of 7.6%.

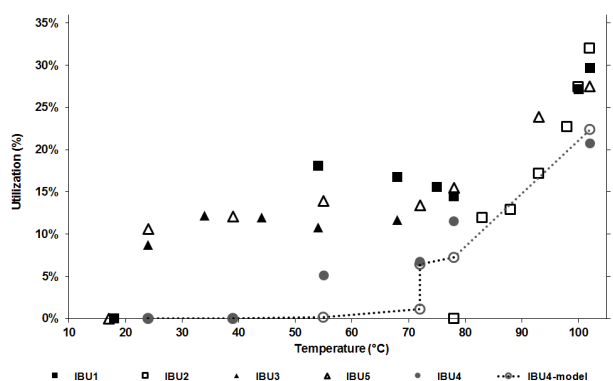

**Figure 5**. Hops α-acids utilization before boiling, as a function of temperature, for selected production tests (IBU1-IBU5). The theoretical curve for the "traditional" test IBU4 is also shown (Malowicki and Shellhammer, 2005).

All tests carried out by means of HC-assisted brewing achieve utilization factors around 30% at boiling inception, a result shared by all the other performed tests, even when using different kinds or combinations of hops (not shown). Few noticeable evidences emerge, showing the advantages arising from the use of HC processes in the brewing practice.

The utilization factors before boiling exceeded the maximum value achieved by the IBU4 test carried out with traditional B-50 equipment by up to more than 1.5 times (32% against 21%). The utilization factors achieved in IBU1, IBU3 and IBU5, when hops were pitched from the very beginning of the respective processes (temperatures around 20°C) grew very quickly to about 12% at 40°C, with a peak of 18% at 54°C, when the utilization factor in IBU4 conventional production test was merely 5%. Over that temperature, the growth of the utilization factors stalled or even reversed (IBU1, for instance), up to 75°C to 80°C, likely due to degradation of iso-α-acids (Lajçi et al., 2013; Malowicki and Shellhammer, 2005), after which it resumed very quickly up to the boiling point. In the IBU2 production, when hops were pitched at 78°C, the utilization factor grew quickly and achieved the highest peak value of around 32% at the boiling point, showing that it is convenient to pitch the hops after the removal of malts.

Moreover, Figure 5 shows that the evolution of the utilization factor in the traditional IBU4 test closely fits the theoretical curve (IBU4-model) (Malowicki and Shellhammer, 2005), making the same test fairly representative of traditional processes.

Finally, the utilization factor in IBU4 test increased, until approaching the values achieved with the new installation, only after boiling for about one hour (not shown). Therefore, for all purposes of net utilization of hops' α-acids, boiling becomes unessential in HC-assisted brewing, regardless of cavitating or caged malts.

The sensitivity of the utilization factors to the cavitation number proved to be very small. For example, production test IBU5 was carried out with an average hydraulic pressure of 1.5 atm, resulting in an average CN twice higher compared with that characterizing the other cavitational tests. Yet, results were barely distinguishable.

In summary, in HC-assisted brewing, hops should be conveniently pitched after the removal of the malts (temperature around 78°C), with no need to apply additional pressure beyond the atmospheric one, while the net utilization of α-acids is practically completed before the onset of bulk boiling, which therefore becomes unessential.

A possible interpretation for the above results is the following. Cavitation mimics boiling although by means of depressurization instead of heating (indeed, the liquid medium boils in *vena contracta* downstream the cavitation nozzle, although each cavity survives during very few msec) (Šarc et al., 2016). Moreover, the pulverization process observed for malted barley acts as well upon hops, thereby greatly enhancing the exposed surface of the hops and the mass transfer to the wort. Hence, the hop oil and α-acids extraction can be completed at lower temperatures and in particular without classical bulk boiling.

Although not directly measured in the performed experiments, no flavor attributable to DMS could ever be felt in any of the beers produced by means of HC-assisted brewing. Indeed, hydrodynamic cavitation processes are long known to lead to very effective liquid degassing (Gogate and Pandit, 2011; Iben et al., 2015; Senthil Kumar et al., 2000), so much that it can be hypothesized that undesired volatile aromatic compounds are safely and promptly expelled throughout their generation, as well as the respective concentration increase with rising temperature is prevented. Therefore, also from this point of view, boiling becomes unessential in HC-assisted brewing.

The capability of hydrodynamic cavitation to inactivate pathogen or undesired microorganisms was discussed in Section 1, as well as proved in a previous study of ours to occur at moderate temperatures (lower than 60°C) (Albanese et al., 2015). In fact, correct long-term preservation (more than 9 months at the time of writing) of bottled beers produced by HC-assisted brewing provided a qualitative confirmation.

Finally, foamability and foam stability, the enhancement of the latter being one of the purposes of wort boiling in conventional brewing as stated in this same Section, were verified for test C6 at 124 and 260 days after brewing: Figure 6 shows an acceptable foam stability in time despite a slight reduction. The choice of test C6 was motivated by its longest cavitation process – most of which performed after malts' removal – across all the tests listed in Table 1, yet the same results were observed to hold for all the other tests.

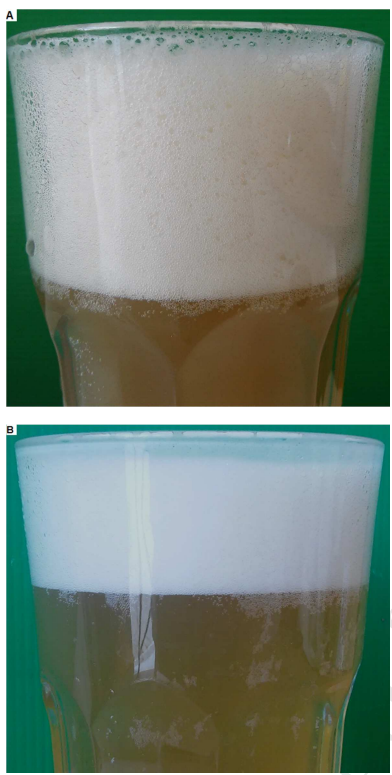

**Figure 6**. Foam stability of the beer obtained in production test C6, 124 days (A) and 260 days (B) after brewing.

**4.6 Energy efficiency**

The primary energy utilization over the whole life cycle of beer production, from barley malting to filling and other processes, has recently been estimated in the range 1-1.7 GJ per hectoliter of the final product (Amienyo and Azapagic, 2016), with typical electricity demand of modern breweries on average around 25-30% of the overall energy requirements (Kubule et al., 2016). More in detail, such energy demand includes 8-16 kWh of electricity and 150-180 MJ thermal (fuel) energy per hectoliter of beer produced (Olajire, 2012; Sturm et al., 2013), with energy costs amounting up to 8-9% of breweries' overall costs (Kubule et al., 2016; Sturm et al., 2013). About 60% of thermal energy (i.e. around 100 MJ/hL) is involved in wort heating and boiling (including mashing and hopping), while estimates are more uncertain for electricity, which is used for numerous tasks including room air conditioning, lighting and so on (Olajire, 2012). A conservative estimate of 30% average electricity consumption for wort heating and boiling amounts to 4 kWh/hL per hectoliter. It is worth noting that such a figure pertains to medium to large breweries (at least one million hectoliters per year), which are far more energy efficient than small and craft brewing plants (Muster-Slawitsch et al., 2011; Sturm et al., 2013). Moreover, due to the moderate temperatures involved, the thermal and electric energy demand can be straightforwardly added up, totaling about 32 kWh/hL (Muster-Slawitsch et al., 2011).

HC-assisted brewing, contrary to traditional one, makes boiling an unnecessary step. Overall, the electricity consumption in test CO1, carried out with cavitating malts (Table 1), was about 21 kWh/hL from the switching-on of the equipment to the wort delivery to the fermenter (mashing and hopping stages). When compared with the above-mentioned conservative estimate of 32 kWh/hL for traditional medium to large breweries, the energy saving amounts to about 34%, let alone the fact that – as mentioned above – craft breweries such as those emerging in the global market, as well as more relevant to the volumes implied by the HC brewing device into consideration, are even less energy efficient.

Indeed, limited to the brewing stages involving water and wort heating (i.e., from water pre-heating to boiling), it was observed that the specific energy saving achieved in test CO1 in comparison to test B1, involving the most similar process among those carried out with the traditional device, was about 33% (specific energy consumption 21 kWh/hL and 31.5 kWh/hL, respectively). Such result was mainly due to a 12% saving before malts removal (water pre-heating mashing stage), as shown in Figure 4(B), and to the absence of the boiling stage in test CO1, the latter stage accounting for about 25% of the overall energy consumption in test B1. Moreover, it should be recalled that the respective brewing yields were very different: as shown in Figure 4(A), the peak starch extraction was 91% in test CO1 and only 60% in test B2.

In addition, it should be considered that operations such as dry milling of malts and cleaning and sanitization of installations are unnecessary or largely reduced, respectively, due to HC processes. Consequently, their respective energy demands, which have been partially unaccounted for in the above, add up to the higher energy efficiency of HC-assisted brewing. However, it should be noted that malt milling, in particular, is a low energy intensity process, requiring just around 2 to 4.4 kWh per metric ton of malt (Frank et al., 2016), that, assuming a malt use around 17 kg/hL as shown in Table 1, translates into a specific energy consumption lower than a mere 0.08 kWh/hL. Rather, what is relevant to the removal of the dry milling stage by means of HC-assisted brewing are the savings in the cost of equipment and in the processing time.

Moreover, the pulverization of spent malts shown in Section 4.2, leading to the improvement of their biodegradability, as was recently proven (Montusiewicz et al., 2016), could lead to the respective energy recovery in the form of biogas, thereby adding to the overall energy efficiency of HC-assisted brewing.

It may also be noted that we did not undertake any passive energy efficiency measure, such as for example thermal insulation of the external metal surfaces of pipes and tank, resulting in a substantial heat dissipation (Baurov et al., 2014), which would be easily reduced in an industrial setup, further adding to the energy performance of HC-assisted brewing.

Finally, the energy efficient HC-assisted brewing is susceptible to be powered by solar thermal and photovoltaic energy with even greater ease compared to conventional production processes (Mauthner et al., 2014; Muster-Slawitsch et al., 2011).

## 4.7 Safety and reliability

Hydrodynamic cavitation has long been regarded as a harmful phenomenon for even the hardest material structures such as impellers and stainless steel surfaces, mainly as a result of high pressure wave emission (Fortes Patella and Reboud, 1998). However, Figure 7 shows no damage to the internal surface bounding the cavitation area (downstream of the nozzle) of the Venturi tube reactor used in the tests listed in Table 1, as well as in several other production sessions, totaling about one thousand hours of operation. The weldings, visible at the top and the bottom and representing the weaker part of the surface, were intact too. Apparently, the strict adherence of the geometry of the used cavitation reactor with recently published stringent recommendations has produced this highly desirable absence of any structural damage (Šarc et al., 2017).

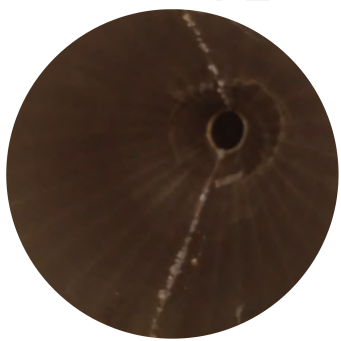

**Figure 7**. Internal surface of the Venturi-tube cavitational reactor, downstream the nozzle, after about 1,000 hours of operation.

About possible harmful effects on the final product, a potential issue raised by the use of hydrodynamic cavitation processes with liquid foods is oxidation, as mentioned in Section 1, where it was argued that – within the range of cavitation regimes established in such applications – oxidative effects are generally negligible. That argument was confirmed by the observation of the beers produced after HC-assisted brewing, whose colors strictly matched the expectations based on their respective recipes while showing no sign of oxidation. For example, beer from test C6, which involved the longest cavitation process across all the tests listed in Table 1, 16 days after the test date showed a color level 2.8 on the SRM scale and 6 on the EBC scale, as well as an alcohol by volume 5%, both matching a Blonde Ale beer style as expected from its recipe (Stewart and Priest, 2006). The color scale became 2.5 (SRM) and 5 (EBC) 260 days after brewing (see also Figure 6(B) for a visual insight), showing a slight clarification as expected from the normal evolution of bottled beer. Although subjective and not assessed so far by an independent panel, the aroma, taste and flavor of all the produced beers after HC-assisted brewing looked like to match the expectations from their respective recipes.

## 5. Discussion

This study was aimed at filling the gap left by HC applications to liquid foods other than beer, as discussed in Section 3, as well as by the allegedly only study dealing with HC-assisted beer brewing (Safonova et al., 2015), that, as discussed in Section 3, involved very small laboratory-scale volumes as well as a rotating-pulsating cavitator whose upward scalability was not proven. The results of this study were achieved at the scale of a real micro-brewery (gross volume capacity 230 L) and using a circular Venturi-tube reactor whose geometry adheres to recent stringent recommendations (Albanese et al., 2015; Šarc et al., 2017), as well as were compared with the outcomes of traditional brewing processes based on data drawn from both the available literature and direct experiments.

The observed dramatic decrease of saccharification temperature under moderate hydrodynamic cavitation, in particular, was a very relevant achievement (Section 4.3), which closely agreed with previous results discussed in Section 3 (Knorr et al., 2011; Sinisterra, 1992).

Equally important process improvements concerned the increase in starch extraction efficiency and the dramatic reduction of the time to peak extraction, as discussed in Section 4.4, as well as the completion of the extraction of hops' $\alpha$-acids before boiling, shown in Section 4.5 along with its possible interpretation. Consequently, under HC-assisted brewing, the beer wort can be conveyed to the fermenters after being heated up to about 100°C, without any boiling, thus saving process time and energy (Section 4.6).

All the above, while retaining safety and reliability (Section 4.7), scalability, virtually universal application to any brewing recipe, beer quality.

Figure 8 shows a simplified scheme of traditional brewing stages from malt (or grains) milling down to fermentation, next to the same scheme related to HC-assisted brewing, as well as to a list of the respective estimated energy savings based on literature and observations discussed in this study. The overall energy saving achievable by means of HC-assisted brewing from water pre-heating to boiling is estimated at the level of 30% or greater.

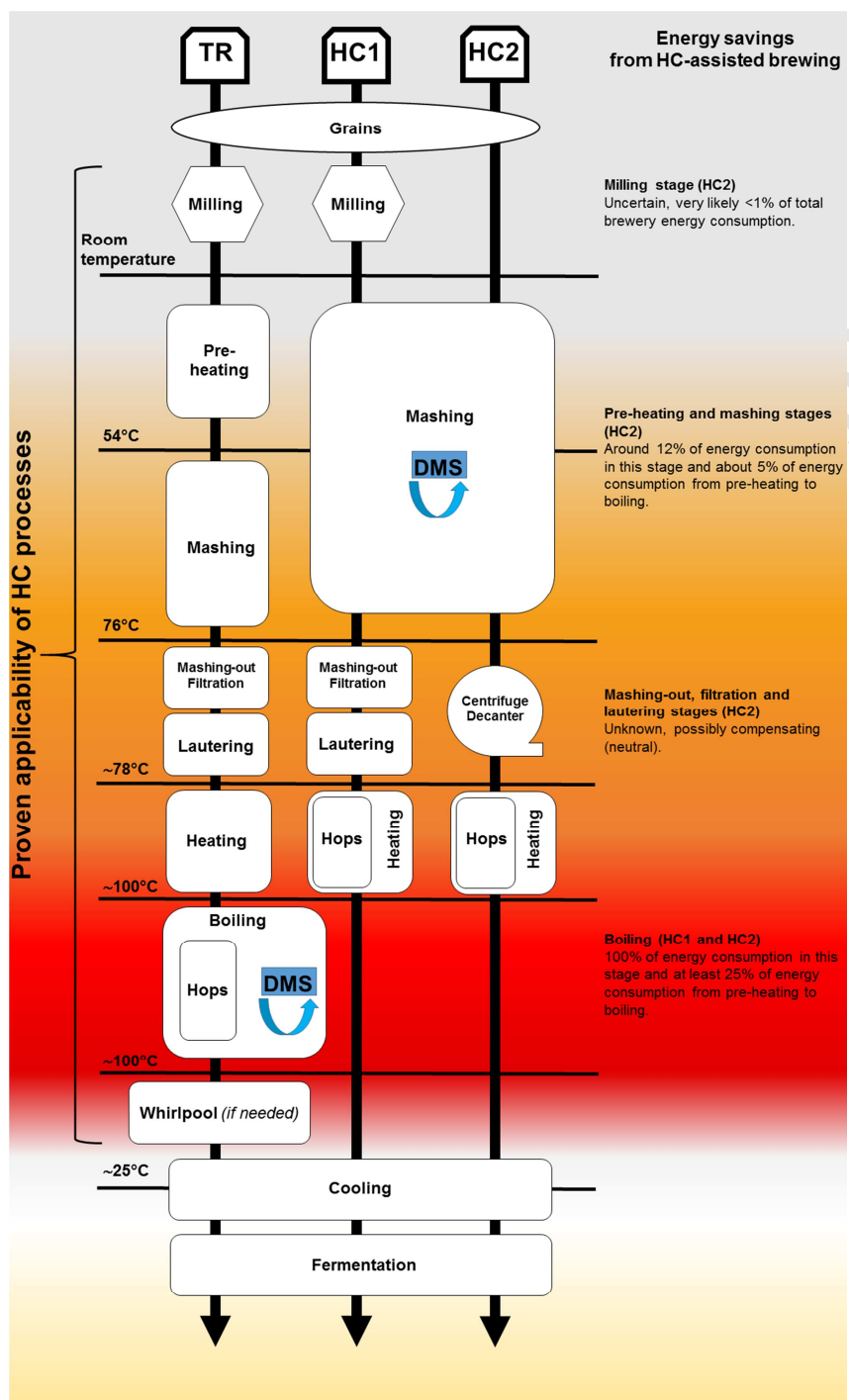

**Figure 8**. Simplified scheme of brewing stages for traditional (TR) and HC-assisted (HC1 – BIAB setup, HC2 – cavitating malts setup) processes, along with the respective energy saving estimates.

The importance of these results arises from their novelty – the successful, advantageous, safe and reliable application of HC processes in the brewing industry whose equipment and basic practices have barely changed over time – as well as from the impressive and growing size of the respective market, skyrocketed to over 200 billion liters per year.

The authors acknowledge that the experimental part of this study is not exempt from limitations, the most important of which are the following. First, further experiments should be performed, especially as duplicates or more of some of those listed in Table 1, in order to provide a more meaningful statistics, even if few experiments used at least the same ingredients and all the HC-assisted brewing tests used the same equipment.

Second, especially important issues such as water disinfection from pathogen microorganisms and remediation from organic pollutants, that hydrodynamic cavitation has been proved able to tackle, were discussed only based on the available literature, among which a previous work from the authors of this study (Albanese et al., 2015), the latter however dealing with the inactivation of a single yeast strain. Recently, very important findings and recommendations were produced, involving an extension of the cavitation regimes especially for water disinfection, which should be included in future experiments (Dular et al., 2016). The same holds for the early expulsion of undesired volatile aromatic compounds such as DMS (whose flavor, anyway, was clearly absent in the produced beers), which should be measured directly in future experiments.

Third, only one HC setup was used, using a circular Venturi-tube reactor. However, recent evidence suggests that slit Venturi reactors outperform circular ones as well as other setups, especially in terms of cavitational yield, as defined in Section 1.

Fourth, independent panels should assess the quality of the beers produced by HC-assisted brewing.

## 6. Conclusions

By means of theory and experiments, this study has shown that the emerging technology of controlled hydrodynamic cavitation has the potential to be successfully introduced in the field of beer-brewing at the industrial scale, leading to significant advantages without apparent drawbacks.

Dramatic reduction of saccharification temperature by 35°C, increased and accelerated peak starch extraction (up to 30% higher extraction efficiency with 12% less consumed energy), significant reduction of process time after traditional stages such as dry milling and boiling are made unessential (up to 120 min only for the boiling stage), represent the most important benefits to the brewing process. Other major advantages concern energy saving, estimated at the level of 30% or greater, shorter cleaning time and overall simplification of both structural setup and operational management of brewing processes.

Despite few acknowledged limitations, the discussed innovation in beer-brewing by means of controlled hydrodynamic cavitation, which was proved on a real microbrewery scale, can be far-reaching and able to quickly spread across the respective large, growing (more than 200 billion liters per year) and increasingly specialized industry, potentially changing the chemistry, the engineering, and the environmental footprint of the processes and therefore producing a major impact.

**Acknowledgements**

L.A. and F.M. were partially funded by Tuscany regional Government under the project T.I.L.A. (Innovative Technology for Liquid Foods, Grant N°. 0001276 signed on April 30, 2014). The research was carried out under a cooperation between CNR-IBIMET and the company Bysea S.r.l., with joint patent submitted on August 9, 2016, international application No. PCT/IT/2016/000194 "A method and relative apparatus for the production of beer", pending. F. Martelli and C. Capriolo are gratefully acknowledged for continuous support in project management and technical help, respectively.

**Declaration of interest**

L.A. and F.M. were appointed as Inventors in the patent submitted on August 9, 2016, international application No. PCT/IT/2016/000194 "A method and relative apparatus for the production of beer", pending.

# 5 References

Abid, M., Jabbar, S., Wu, T., Hashim, M.M., Hu, B., Lei, S., Zeng, X., 2014. Sonication enhances polyphenolic compounds, sugars, carotenoids and mineral elements of apple juice. Ultrason. Sonochem. 21, 93–7. doi:10.1016/j.ultsonch.2013.06.002

Abid, M., Jabbar, S., Wu, T., Hashim, M.M., Hu, B., Lei, S., Zhang, X., Zeng, X., 2013. Effect of ultrasound on different quality parameters of apple juice. Ultrason. Sonochem. 20, 1182–7. doi:10.1016/j.ultsonch.2013.02.010

Ahmed, J., Thomas, L., Arfat, Y.A., 2016. Effects of high hydrostatic pressure on functional, thermal, rheological and structural properties of β-D-glucan concentrate dough. LWT - Food Sci. Technol. 70, 63–70. doi:10.1016/j.lwt.2016.02.027

Albanese, L., Ciriminna, R., Meneguzzo, F., Pagliaro, M., 2015. Energy efficient inactivation of Saccharomyces cerevisiae via controlled hydrodynamic cavitation. Energy Sci. Eng. 3, 221–238. doi:10.1002/ese3.62

Ambrosi, A., Cardozo, N.S.M., Tessaro, I.C., 2014. Membrane Separation Processes for the Beer Industry: a Review and State of the Art. Food Bioprocess Technol. 7, 921–936. doi:10.1007/s11947-014-1275-0

Amienyo, D., Azapagic, A., 2016. Life cycle environmental impacts and costs of beer production and consumption in the UK. Int. J. Life Cycle Assess. 21, 492–509. doi:10.1007/s11367-016-1028-6

Bagal, M. V, Gogate, P.R., 2014. Wastewater treatment using hybrid treatment schemes based on cavitation and Fenton chemistry: a review. Ultrason. Sonochem. 21, 1–14. doi:10.1016/j.ultsonch.2013.07.009

Batoeva, A.A., Aseev, D.G., Sizykh, M.R., Vol'nov, I.N., 2011. A study of hydrodynamic cavitation generated by low pressure jet devices. Russ. J. Appl. Chem. 84, 1366–1370. doi:10.1134/S107042721108012X

Baurov, Y.A., Albanese, L., Meneguzzo, F., 2014. New force and new heat. Am. J. Astron. Astrophys. 2, 47–53. doi:10.11648/j.ajaa.s.20140202.17

Bohačenko, I., Chmelík, J., Psota, V., 2006. Determination of the contents of A- and B-starches in barley using Low Angle Laser Light Scattering. Czech J. Food Sci. 24, 11–18.

Bokulich, N.A., Bamforth, C.W., 2013. The microbiology of malting and brewing. Microbiol. Mol. Biol. Rev. 77, 157–72. doi:10.1128/MMBR.00060-12

Choi, E.J., Ahn, H., Kim, M., Han, H., Kim, W.J., 2015. Effect of ultrasonication on fermentation kinetics of beer using six-row barley cultivated in Korea. J. Inst. Brew. 121, 510–517. doi:10.1002/jib.262

Choi, J.-H., Kang, J.-W., Mijanur Rahman, A.T.M., Lee, S.J., 2016. Increasing fermentable

sugar yields by high-pressure treatment during beer mashing. J. Inst. Brew. 122, 143–146. doi:10.1002/jib.285

Ciriminna, R., Albanese, L., Meneguzzo, F., Pagliaro, M., 2016a. Wastewater remediation via controlled hydrocavitation. Environ. Rev. doi:10.1139/er-2016-0064

Ciriminna, R., Albanese, L., Meneguzzo, F., Pagliaro, M., 2016b. Hydrogen Peroxide: A Key Chemical for Tomorrow's Sustainable Development. ChemSusChem. doi:10.1002/cssc.201600895

Dindar, E., 2016. An Overview of the Application of Hydrodinamic Cavitation for the Intensification of Wastewater Treatment Applications: A Review. Innov. Energy Res. 5, 1–7. doi:10.4172/ier.1000137

Dular, M., Griessler-Bulc, T., Gutierrez-Aguirre, I., Heath, E., Kosjek, T., Krivograd Klemenčič, A., Oder, M., Petkovšek, M., Rački, N., Ravnikar, M., Šarc, A., Širok, B., Zupanc, M., Žitnik, M., Kompare, B., 2016. Use of hydrodynamic cavitation in (waste)water treatment. Ultrason. Sonochem. 29, 577–588. doi:10.1016/j.ultsonch.2015.10.010

Fortes Patella, R., Reboud, J.-L., 1998. A New Approach to Evaluate the Cavitation Erosion Power. ASME J. Fluids Eng. 120, 335–344. doi:10.1115/1.2820653

Frank, A., Scheel, R., Ag, B., 2016. Optimized Dry Processing Using the Newest Generation of Grinding Equipment. MBAA Tech. Q. 53, 97–102. doi:10.1094/TQ-53-2-0502-01

Gogate, P.R., 2011. Hydrodynamic Cavitation for Food and Water Processing. Food Bioprocess Technol. 4, 996–1011. doi:10.1007/s11947-010-0418-1

Gogate, P.R., 2002. Cavitation: an auxiliary technique in wastewater treatment schemes. Adv. Environ. Res. 6, 335–358. doi:10.1016/S1093-0191(01)00067-3

Gogate, P.R., Pandit, A.B., 2011. Cavitation Generation and Usage Without Ultrasound: Hydrodynamic Cavitation, in: Pankaj, D.S., Ashokkumar, M. (Eds.), Theoretical and Experimental Sonochemistry Involving Inorganic Systems. Springer Netherlands, Dordrecht, pp. 69–106. doi:10.1007/978-90-481-3887-6

Gogate, P.R., Pandit, A.B., 2005. A review and assessment of hydrodynamic cavitation as a technology for the future. Ultrason. Sonochem. 12, 21–27. doi:10.1016/j.ultsonch.2004.03.007

Gogate, P.R., Pandit, A.B., 2001. Hydrodynamic cavitation reactors: a state of the art review. Rev. Chem. Eng. 17, 1–85. doi:10.1515/REVCE.2001.17.1.1

Gogate, P.R., Shirgaonkar, I.Z., Sivakumar, M., Senthilkumar, P., Vichare, N.P., Pandit, A.B., 2001. Cavitation reactors: Efficiency assessment using a model reaction. AIChE J. 47, 2526–2538. doi:10.1002/aic.690471115

Hager, A.-S., Taylor, J.P., Waters, D.M., Arendt, E.K., 2014. Gluten free beer – A review. Trends Food Sci. Technol. 36, 44–54. doi:10.1016/j.tifs.2014.01.001

He, B., Zhang, L.-L., Yue, X.-Y., Liang, J., Jiang, J., Gao, X.-L., Yue, P.-X., 2016. Optimization of Ultrasound-Assisted Extraction of phenolic compounds and anthocyanins from blueberry (Vaccinium ashei) wine pomace. Food Chem. 204, 70–76. doi:10.1016/j.foodchem.2016.02.094

Iben, U., Wolf, F., Freudigmann, H.-A., Fröhlich, J., Heller, W., 2015. Optical measurements of gas bubbles in oil behind a cavitating micro-orifice flow. Exp. Fluids 56, 114. doi:10.1007/s00348-015-1979-6

Knorr, D., Froehling, A., Jaeger, H., Reineke, K., Schlueter, O., Schoessler, K., 2011. Emerging technologies in food processing. Annu. Rev. Food Sci. Technol. 2, 203–35. doi:10.1146/annurev.food.102308.124129

Kubule, A., Zogla, L., Ikaunieks, J., Rosa, M., 2016. Highlights on energy efficiency improvements: A case of a small brewery. J. Clean. Prod. 138, 275–286. doi:10.1016/j.jclepro.2016.02.131

Lajçi, X., Dodbiba, P., Lajçi, N., 2013. The Evaluation of Humulus Lupulus Utilisation During Wort Boiling and in Beer, in: The 1st International Conference on Research and Education – Challenges Toward the Future (ICRAE2013). University of Shkodra "Luigj Gurakuqi", Shkodra, Albania, pp. 1–8.

Langone, M., Ferrentino, R., Trombino, G., Waubert De Puiseau, D., Andreottola, G., Rada, E.C., Ragazzi, M., 2015. Application of a novel hydrodynamic cavitation system in wastewater treatment plants. UPB Sci. Bull. Ser. D Mech. Eng. 77, 225–234.

Lee, I., Han, J.-I., 2013. The effects of waste-activated sludge pretreatment using hydrodynamic cavitation for methane production. Ultrason. Sonochem. 20, 1450–1455. doi:10.1016/j.ultsonch.2013.03.006

Maeng, J.W., Lee, E.Y., Bae, J.H., 2010. Optimization of the Hydrodynamic Sludge Pre-Treatment System with Venturi Tubes. Water Pract. Technol. 5, 1–10. doi:10.2166/wpt.2010.034

Malowicki, M.G., Shellhammer, T.H., 2005. Isomerization and degradation kinetics of hop (Humulus lupulus) acids in a model wort-boiling system. J. Agric. Food Chem. 53, 4434–4439. doi:10.1021/jf0481296

Matsuura, K., Hirotsune, M., Nunokawa, Y., Satoh, M., Honda, K., 1994. Acceleration of cell growth and ester formation by ultrasonic wave irradiation. J. Ferment. Bioeng. 77, 36–40. doi:10.1016/0922-338X(94)90205-4

Mauthner, F., Hubmann, M., Brunner, C., Fink, C., 2014. Manufacture of Malt and Beer with

Low Temperature Solar Process Heat. Energy Procedia 48, 1188–1193. doi:10.1016/j.egypro.2014.02.134

Milani, E.A., Ramsey, J.G., Silva, F.V.M., 2016. High pressure processing and thermosonication of beer: Comparing the energy requirements and Saccharomyces cerevisiae ascospores inactivation with thermal processing and modeling. J. Food Eng. 181, 35–41. doi:10.1016/j.jfoodeng.2016.02.023

Milly, P.J., Toledo, R.T., Harrison, M. a, Armstead, D., 2007. Inactivation of food spoilage microorganisms by hydrodynamic cavitation to achieve pasteurization and sterilization of fluid foods. J. Food Sci. 72, M414-22. doi:10.1111/j.1750-3841.2007.00543.x

Milly, P.J., Toledo, R.T., Kerr, W.L., Armstead, D., 2008. Hydrodynamic Cavitation: Characterization of a Novel Design with Energy Considerations for the Inactivation of Saccharomyces cerevisiae in Apple Juice. J. Food Sci. 73, M298–M303. doi:10.1111/j.1750-3841.2008.00827.x

Montusiewicz, A., Pasieczna-Patkowska, S., Lebiocka, M., Szaja, A., Szymańska-Chargot, M., 2016. Hydrodynamic cavitation of brewery spent grain diluted by wastewater. Chem. Eng. J. 1–11. doi:10.1016/j.cej.2016.10.132

Muster-Slawitsch, B., Weiss, W., Schnitzer, H., Brunner, C., 2011. The green brewery concept – Energy efficiency and the use of renewable energy sources in breweries. Appl. Therm. Eng. 31, 2123–2134. doi:10.1016/j.applthermaleng.2011.03.033

Ngadi, M.O., Latheef, M. Bin, Kassama, L., 2012. Emerging technologies for microbial control in food processing, in: Boye, J.I., Arcand, Y. (Eds.), Green Technologies in Food Production and Processing, Food Engineering Series. Springer US, Boston, MA, pp. 363–411. doi:10.1007/978-1-4614-1587-9_14

Olajire, A.A., 2012. The brewing industry and environmental challenges. J. Clean. Prod. 1–21. doi:10.1016/j.jclepro.2012.03.003

Patil, P.N., Gogate, P.R., Csoka, L., Dregelyi-Kiss, A., Horvath, M., 2016. Intensification of biogas production using pretreatment based on hydrodynamic cavitation. Ultrason. Sonochem. 30, 79–86. doi:10.1016/j.ultsonch.2015.11.009

Pettigrew, L., Blomenhofer, V., Hubert, S., Groß, F., Delgado, A., 2015. Optimisation of water usage in a brewery clean-in-place system using reference nets. J. Clean. Prod. 87, 583–593. doi:10.1016/j.jclepro.2014.10.072

Pires, E., Brányik, T., 2015. Biochemistry of Beer Fermentation. Springer International Publishing AG Switzerland. doi:10.1007/978-3-319-15189-2

Rajoriya, S., Carpenter, J., Saharan, V.K., Pandit, A.B., 2016. Hydrodynamic cavitation: an advanced oxidation process for the degradation of bio-refractory pollutants. Rev. Chem.

Eng. doi:10.1515/revce-2015-0075

Safonova, E.A., Potapov, A.N., Vagaytseva, E.A., 2015. Intensification of technological processes of beer production using rotary-pulsation apparatus. Food Process. Tech. Technol. 36, 74–81 (in Russian).

Šarc, A., Oder, M., Dular, M., 2016. Can rapid pressure decrease induced by supercavitation efficiently eradicate Legionella pneumophilabacteria? Desalin. Water Treat. 57, 2184–2194. doi:10.1080/19443994.2014.979240

Šarc, A., Stepišnik-perdih, T., Petkovšek, M., Dular, M., 2017. The issue of cavitation number value in studies of water treatment by. Ultrason. Sonochem. 34, 51–59. doi:10.1016/j.ultsonch.2016.05.020

Scheuren, H., Baldus, M., Methner, F.-J., Dillenburger, M., 2016. Evaporation behaviour of DMS in an aqueous solution at infinite dilution - a review. J. Inst. Brew. 122, 181–190. doi:10.1002/jib.301

Scheuren, H., Dillenburger, M., Methner, F.-J., 2015. A new proposal for the quantification of homogeneity in the wort boiling process. J. Inst. Brew. 121, 204–206. doi:10.1002/jib.222

Scheuren, H., Sommer, K., Dillenburger, M., 2015. Explanation for the increase in free dimethyl sulphide during mashing. J. Inst. Brew. 121, 418–420. doi:10.1002/jib.234

Scheuren, H., Tippmann, J., Methner, F.J., Sommer, K., 2014. Decomposition kinetics of dimethyl sulphide. J. Inst. Brew. 120, 474–476. doi:10.1002/jib.156

Senthil Kumar, P., Siva Kumar, M., Pandit, A.B., 2000. Experimental quantification of chemical effects of hydrodynamic cavitation. Chem. Eng. Sci. 55, 1633–1639. doi:10.1016/S0009-2509(99)00435-2

Serie e-SH (in Italian) [WWW Document], 2016. . Lowara - a xylem Brand. URL http://www.xylemitalia.it/Prodotti/images/pdf/eSH_st.pdf (accessed 11.19.16).

Sfakianakis, P., Tzia, C., 2014. Conventional and Innovative Processing of Milk for Yogurt Manufacture; Development of Texture and Flavor: A Review. Foods 3, 176–193. doi:10.3390/foods3010176

Sinisterra, J.V., 1992. Application of ultrasound to biotechnology: an overview. Ultrasonics 30, 180–185. doi:10.1016/0041-624X(92)90070-3

Soyama, H., Hoshino, J., 2016. Enhancing the aggressive intensity of hydrodynamic cavitation through a Venturi tube by increasing the pressure in the region where the bubbles collapse. AIP Adv. 6, 45113. doi:10.1063/1.4947572

Stack, M., Gartland, M., Keane, T., 2016. Path Dependency, Behavioral Lock-in and the International Market for Beer, in: Brewing, Beer and Pubs. Palgrave Macmillan UK, London, pp. 54–73. doi:10.1057/9781137466181_4

Stewart, G.G., Priest, F.G., 2006. Handbook of Brewing, Second Edi. ed. CRC Press, Boca Raton (FL, USA).

Sturm, B., Hugenschmidt, S., Joyce, S., Hofacker, W., Roskilly, A.P., 2013. Opportunities and barriers for efficient energy use in a medium-sized brewery. Appl. Therm. Eng. 53, 397–404. doi:10.1016/j.applthermaleng.2012.05.006

Szwajgier, D., 2011. Dry and Wet Milling of Malt . A Preliminary Study Comparing Fermentable Sugar , Total Protein , Total Phenolics and the Ferulic Acid Content in Non-Hopped Worts. J. Inst. Brew. 117, 569–577. doi:10.1002/j.2050-0416.2011.tb00505.x

Tao, Y., Cai, J., Huai, X., Liu, B., Guo, Z., 2016. Application of hydrodynamic cavitation into wastewater treatment: A review. Chem. Eng. Technol. 1363–1376. doi:10.1002/ceat.201500362

Yasui, K., Tuziuti, T., Sivakumar, M., Iida, Y., 2004. Sonoluminescence. Appl. Spectrosc. Rev. 39, 399–436. doi:10.1081/ASR-200030202

Yusaf, T., Al-Juboori, R. a., 2014. Alternative methods of microorganism disruption for agricultural applications. Appl. Energy 114, 909–923. doi:10.1016/j.apenergy.2013.08.085

**Beer-brewing powered by controlled hydrodynamic cavitation: theory and real-scale experiments**

**Highlights**

- First ever real-scale application of hydrodynamic cavitation to beer-brewing

- Controlled hydrodynamic cavitation assisted brewing is safe, reliable, scalable

- The application allows large savings of process time, energy and washwater

- Traditional malts dry milling and wort boiling become unessential

- Accelerated saccharification, starch extraction and utilization of hops α-acids